\documentclass[pre,showpacs,twocolumn]{revtex4}
\usepackage{epsfig}
\usepackage{dcolumn}
\usepackage{bm}

\newcommand{\be}{\begin{equation}}
\newcommand{\ee}{\end{equation}}
\newcommand{\bc}{\begin{center}}
\newcommand{\ec}{\end{center}}
\newcommand{\bi}{\begin{itemize}}
\newcommand{\ei}{\end{itemize}}
\newcommand{\ba}{\begin{eqnarray}}
\newcommand{\ea}{\end{eqnarray}}

\newcommand{\ignore}[1]{}
\newcommand{\mean}[1]{\left\langle #1 \right\rangle}
\newcommand{\abs}[1]{\left| #1 \right|}

\begin{document}
\title{First order phase transition in Ising model on two connected Barabasi-Albert networks.}
\author{Krzysztof Suchecki}
\email{suchecki@if.pw.edu.pl} \affiliation{Faculty of Physics, Center of Excellence for Complex Systems Research \\Warsaw University of Technology \\ Koszykowa 75, PL--00-662 Warsaw, Poland}
\author{Janusz A. Ho{\l}yst}
\email{jholyst@if.pw.edu.pl} \affiliation{Faculty of Physics, Center of Excellence for Complex Systems Research \\Warsaw University of Technology \\ Koszykowa 75, PL--00-662 Warsaw, Poland}
\date{\today}

\begin{abstract}
We investigate the behavior of the Ising model on two connected Barbasi-Albert scale-free networks. We extend previous analysis and show that a first order temperature-driven phase transition occurs in such system. The transition between antiparalelly ordered networks to paralelly ordered networks is shown to be discontinuous. We calculate the critical temperature. We confirm the calculations with numeric simulations using Monte-Carlo methods.

\end{abstract}
\maketitle

\section{Introduction}
Phase transitions are one of most interesting phenomena. While the behavior of systems in noncritical regions may be also interesting, most crucial changes appear in critical regions. It is therefore important to know when such transitions occur, and how do they occur.\\
It is widely known that the classic Ising model displays a second order temperature driven phase transition. The model is throughly investigated, but only on regular lattices. Along with the emergence of complex networks science starting with breakthrough Barabasi-Albert's paper \cite{barabasi}, came study of Ising model in such systems \cite{staufer,bianconi, Dorog,critical,herrero}. Many aspects of the model have been studied, from simple antiferromagnetic interactions and spin-glasses \cite{antiferro,spinglass} to the directed structure of the network \cite{directed}.\\
In our previous work \cite{isingconnect} we have investigated the model on a pair of connected networks. Recent research indicates that one of two phase transitions in such a system is in fact a first order phase transition, not second order like was thought before.\\
In this paper, we investigate the phase transition in a pair of connected networks, show evidence that it is in fact of first order, and back up our analytical calculations with numerical simulations.

\section{Model}
In our study, we consider two interconnected Barabasi-Albert (B-A) networks, where at each node we place an Ising spin. The interactions between the spins are ferromagnetic only.

\begin{figure}[tb]
 \vskip 0.5cm
 \centerline{\epsfig{file=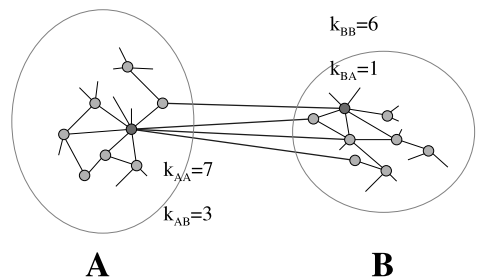,width=.8\columnwidth}}
 \caption{\label{rys_sieci}Two connected B-A networks. A few nodes from each network are shown. The intra-network degrees $k_{AA}$ and $k_{BB}$ as well as inter-network degrees $k_{AB}$ and $k_{BB}$ for two sample nodes are presented.}
\end{figure}

The B-A model is a model of a growing network \cite{barabasi}.
To obtain such a network, one starts with $m$ fully connected nodes, and adds new nodes to the network. Each new node creates $m$ connections to the existing network. The probability that a
connection will be made to a node $i$ is proportional to its degree $k_{i}$. This results in a scale-free network, with a degree distribution $P(k)\sim k^{-3}$.\\
Our two B-A networks are interconnected by $E_{AB}$ links (Fig.\ref{rys_sieci}). Each of these links connects a node in network $A$ with a node in network $B$.
The nodes to be connected are chosen preferentially, i.e. the probability to pick a given node $i$ equals $\Pi_{Ai}=k_{AAi}/\sum_j k_{AAj}$. If we perform linking in this way, the inter-network degree $k_{ABi}$ of a node is statistically proportional its to intra-network degree $k_{AAi}$. \label{pa}

\section{Phase transitions}
The problem of the Ising model on coupled B-A networks has been considered before \cite{isingconnect}. In connected B-A networks, Ising model is characterized by two phase transition in two different critical temperatures $T_{c-}$ and $T_{c+}$. Below $T_{c-}$ there are two possible phases: both networks ordered in with same spin and both networks ordered with opposite spins. At critical temperature $T_{c-}$ the state with antiparallel spin ordering disappears, and between $T_{c-}$ and $T_{c+}$ the system orders only parallely. At $T_{c+}$ and above the temperature is too high for network to remain ordered and it assumes paramagnetic state.\\
As in regular Ising model, the transition at $T_{c+}$ is second order phase transition. However, unlike previous research indicated \cite{isingconnect} the transition at $T_{c-}$ turns out to be of first order.\\
We have performed analytic calculations, numeric map iterations and Monte-Carlo simulations.

\section{Analytic approach}

\begin{figure}[tb]
 \vskip 0.5cm
 \centerline{\epsfig{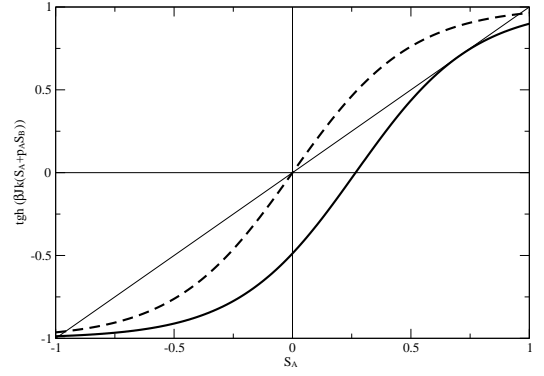}}
 \caption{\label{tanh}Hyperbolic tangent plot. Dashed line is Eq.\ref{map1} for $T<.T_c$, while the solid line is for $T=T_c$. Thin lines are axes and $y=x$ line.}
\end{figure}

We use a mean-field approach to the problem of Ising model. In such approach the self-consistent equation for the average spin
\be
\mean{s_i}=\tanh\left(\beta \sum_j J_{ij} \mean{s_j} + \beta h_i \right)
\ee
can be rewritten as
\be
\mean{s_i}=\tanh\left(\beta J \sum_j \left( \frac{k_i k_j}{E} \mean{s_j} \right) + \beta h_i \right)
\ee
where $\beta=1/T$, the temperature $T$ is measured in units of inverse Boltzmann constant $1/k_B$, averaging is over the canonical ensemble and $h_i$ is the external field acting on node $i$.\\
If we consider two networks that interact, we can treat the influence of the second network as external field $h_i$. Since the inter-network links number is proportional to intra-network links
we can write the full set of equations for two networks
\ba
\mean{s_{Ai}}=\tanh\left(\beta J_{AA} \sum_j \left( \frac{k_{Ai} k_{Aj}}{E_A} \mean{s_{Aj}} \right) + \right. \nonumber \\
\left. + \beta J_{BA} \sum_l \frac{k_{ABi} k_{BAl}}{E_{BA}} \mean{s_{Bl}} \right) ,\\
\mean{s_{Bi}}=\tanh\left(\beta J_{BB} \sum_j \left( \frac{k_{Bi} k_{Bj}}{E_B} \mean{s_{Bj}} \right) + \right. \nonumber \\
\left. + \beta J_{AB} \sum_l \frac{k_{BAi} k_{ABl}}{E_{AB}} \mean{s_{Al}} \right) .
\ea
We introduce a weighted average spin $S=1/E \sum_i k_i s_i$, that is an order parameter for the Ising model on random network of nonhomogenous degree distribution. Additionally we put $k_{ABi}=p_A k_{Ai}$, $k_{BAi}=p_B k_{Bi}$ using the fact that inter-network degrees are proportional to intra-network degrees (see Sect.\ref{pa}). We obtain following equations for the weighted average spins
\ba
S_A= \sum_i \frac{k_{Ai}}{E_{A}} \tanh\left(\beta J_{AA} k_{Ai} S_A + \beta J_{BA} k_{ABi} S_B \right) \label{map1},\\
S_B= \sum_i \frac{k_{Bi}}{E_{B}} \tanh\left(\beta J_{BB} k_{Bi} S_B + \beta J_{AB} k_{BAi} S_A \right) \label{map2}.
\ea
In such a system, a possibility of a first order phase transition exists.\\
Let us consider a pair of random networks of the same size, the same link density and $k=$ const. The right side of the Equation \ref{map1} is hyperbolic tangent, shifted by the value $H=J k_A p_A S_B$ along the $x$ axis. When the temperature $T$ is low, $\beta$ is high and the tangent has three solutions. If temperature increases to critical $T_c$, the curve becomes tangential to the $y=x$ line (Fig.\ref{tanh}). The value of $H$ decreases, since network B is also less ordered at higher temperature so its influence decreases. Below and at $T_c$ we can tell that $S_B=-S_A$ from the symmetry of the system. At $T_c$, the system is unstable and minimal fluctuation of either $S_A$ or $S_B$ causes system to switch over to parallel state.\\


\begin{figure}[t]
 \vskip 0.5cm
 \centerline{\epsfig{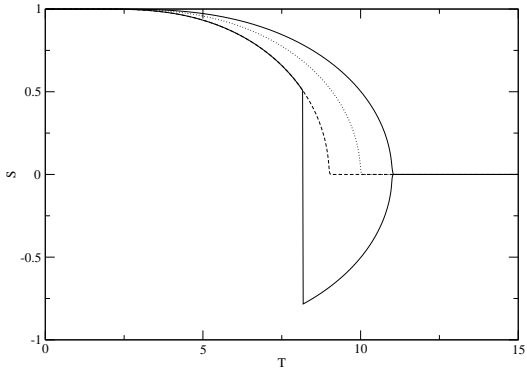}}
 \caption{\label{dyna}Graphs of $S_A(T)$ for $N_A=N_B=5000$ and $k=$ const $=10$. The dotted line is for $p=0$ (unconnected networks) and the rest are for $p=0.1$. The dashed line is the graph with forced $S_A=-S_B$ (this means forced second order phase transition), the solid lines are without such forcing, for parallel and antiparallel initial ordering. The first order phase transition is evident for antiparallel case.}
\end{figure}

At $T_c$, the tangent is tangential to the $y=x$ line. We can write the conditions for $\beta_c$ and $S_{Ac}$
\ba
\frac{\tanh(\beta_c J k_A S_A + \beta_c J k_A p_A S_B)}{S_A}=1 \label{war1} \\
\frac{\partial \tanh(\beta_c J k_A S_A + \beta_c J k_A p_A S_B)}{\partial S_A}=1 \label{war2}
\ea
We have calculated the $\beta_c$ and $S_{Ac}$ from these equations for $S_B=-S_A$ and we obtained
\ba
S_{Ac}=\frac{\ln \left(\sqrt{\beta_c J k_A}+\sqrt{\beta_c J k_A - 1} \right)}{\beta_c J k_A (1-p_A)} \label{solu1} \\
\beta_c = \frac{\ln \frac{1+S_{Ac}}{1-S_{Ac}}}{2 J k_A (1-p_A) S_{Ac}} \label{solu2}
\ea
This set of equations determines the critical point $(T_c,S_{Ac})$ for the first order phase transition between the antiparallel and parallel states.\\
If we multiply Eq.\ref{solu1} by $\beta_c$ and Eq.\ref{solu2} by $S_{Ac}$ we get $\beta_c S_{Ac}$ in both and can compare the right sides, obtaining a relation
\be
S_{Ac}=\frac{\beta_c J k_A +\sqrt{\beta_c J k_A(\beta_c J k_A -1)}-1}{\beta_c J k_A + \sqrt{\beta_c J k_A(\beta_c J k_A -1)}} .
\ee
Comparing this with Eq.\ref{solu1} we obtain a single implicit equation for $\beta_c$ and $p_A$, that can be simplified to get
\ba
p_A = 1 - \frac{1+\sqrt{1-1/(\beta_c J k_A)}}{\beta_c J k_A + \sqrt{\beta_c J k_A (\beta_c J k_A - 1)} -1} \cdot \nonumber \\
\cdot \ln \left( \sqrt{\beta_c J k_A} + \sqrt{\beta_c J k_A -1} \right) \label{pbeta}.
\ea
Drawing $p_A(\beta_c)$ and changing axes yields a dependence of $T_c$ on parameter $p_A$ (see Figure \ref{tc}).\\

We can also approximate the behavior of the solution for small $p_A$. Our conditions (Eq.\ref{war1}-\ref{war2}) can be written
\ba
\tanh(\beta_c J k_A (1-p_A) S_{Ac} )=S_{Ac} \label{war1b}\\
\cosh^2(\beta_c J k_A (1-p_A) S_{Ac} )=\beta_c J k_A \label{war2b}
\ea
If we multiply the equations sidewise, we obtain a single equation for a multiple $X=\beta_c J k_A S_{Ac}$.
\be
\sinh (2 (1-p_A) X)= X
\ee
We know that for very small $p_A$ the value of $S_{Ac}$ is very small, thus $X$ and whole argument of hyperbolic sinus is also small and we can approximate it around $0$
\be
2 (1-p_A) X + (2 (1-p_A) X)^3/6 \approx 2 X
\ee
we can calculate the approximate value of $X$
\be
X \approx \sqrt{\frac{3}{2}p_A}
\ee
Putting the result into the Equation \ref{war2b} we obtain following
\be
\cosh^2 \left((1-p_A)\sqrt{(3/2)p_A}\right)= \beta_c J k_A
\ee
Since the argument of $\cosh^2$ is very small thanks to small $p_A$ value, we can approximate $\cosh^2 x =1+x^2$ and finally obtain
$T_c \approx k_A(1-(3/2) p_A)$

So far, we have concentrated on a case of constant node degree $k$ and two networks of same size. Without such simplifications, the equations are very hard to solve analytically. We have studied more complex cases using map iterations and Monte-Carlo simulations.

\section{Map iterations}

\begin{figure}[t]
 \vskip 0.5cm
 \centerline{\epsfig{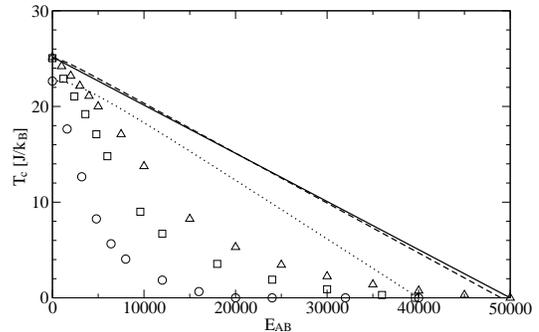}}
 \caption{\label{tode}Dependence of critical temperature $T_c$ on the number of inter-network connections $E_{AB}$ for two Barabasi-Albert networks. Lines are analytic predictions based on second order phase transition assumption, while symbols are critical temperatures obtained form map iterations. Solid line and triangles correspond to $N_A=N_B=5000$, dashed line and squares correspond to $N_A=6000$, $N_B=4000$, while dotted line and circles correspond to $N_A=8000$, $N_B=2000$.}
\end{figure}

\begin{figure}[tb]
 \vskip 0.5cm
 \centerline{\epsfig{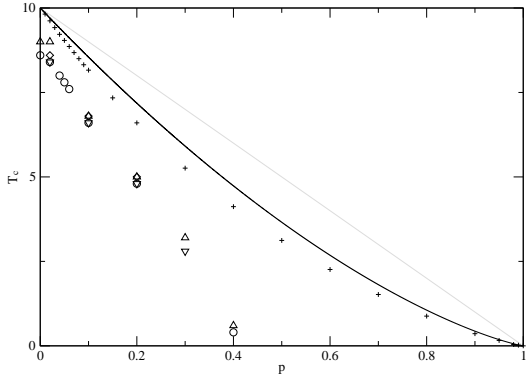}}
 \caption{\label{tc}Dependence of critical temperature $T_c$ on the parameter $p$ for two constant degree networks $k=10$. The straight gray line is $T_c(p)$ if the transition was of second order. The solid line is analytical prediction (Eq.\ref{pbeta}), the plus symbols are map iterations, while circles, triangles and diamonds are results of numerical Monte-Carlo simulations. Circles are for $\tau=100$, triangles up are for $\tau=30$, triangles down are for $\tau=200$. Diamonds are for $\tau=100$ but for networks of size $N=N_A=N_B=50000$.}
\end{figure}

Since the problem of the first order phase transition could not be solved fully analytically, we have used numerical methods.\\
We consider a two-dimensional map
\ba S_A (t+1) = \sum_{k_A} P(k_A) \frac{k_A}{E_A} \tanh\left(\beta J_{AA} k_A S_A (t) + \right. \nonumber \\
\left. + \beta J_{BA} p_A k_A S_B (t) \right) ,\\
S_B (t+1) = \sum_{k_B} P(k_B) \frac{k_B}{E_B} \tanh\left(\beta J_{BB} k_B S_B (t) + \right. \nonumber \\
\left. + \beta J_{AB} p_B k_B S_A (t) \right) .
\ea
where the $S_A(t)$, $S_B(t)$ are variables and the rest are parameters, including given degree distributions $P(k_A)$ and $P(k_B)$. With our definition of weighted spin $S=1/E \sum_i k_i s_i$, where $s_i$ are spin values of node $i$, $k_i$ are degrees and $E$ is number of edges in network, it can have values from range $[-1,1]$. We assume $J_{AA}=J_{BB}=J_{AB}=J_{BA}=J$ and express all temperatures in units of coupling constant $J$ over Boltzmann constant $k$, so we can omit these constants in the equations and have $\beta=1/T$.\\
We investigate the dependence of a stable state spin on the temperature $S(T)$ in the antiparallel ordering of both networks $S_A(T)=1$,$S_B=-1$. Since the system is fully symmetric, below $T_c$ we have $S(T)=S_A(T)=-S_B(T)$. At $T_c$, the systems jumps to the parallel ordering. Due to deterministic nature of these calculations both networks always assume same (negative) spins $S(T)=S_A(T)=S_B(T)<0$. By observing $S(T)$ we can find the critical temperature $T_c$, where a jump between positive and negative spin values occurs (see Fig.\ref{dyna}). Our $S_A(T)=\left( S_A(t_{max}) \right)_{T}$, $S_B(T)=\left(S_B(t_{max}) \right)_{T}$, where $t_{max}=1000$ is the number of iterations of the map that have been performed before we assumed it reached stationary solution for sure.\\
We investigated various $T$ ranges, usually around the critical temperature $T_c$, with the temperature step $\Delta T=0.2$. Our networks were of size $N_A=N_B=5000$ and usually possessed a power law degree distribution $P(k_A)=P(k_B)$ taken from a Barabasi-Albert network growth simulation or constant degree $k=$ const for testing of the analytical equations. Since the networks are same $p_A=p_B=p$.\\

We have investigated the dependence of the critical temperature $T_c$ for two networks with constant $k=10$. The results are in Figure \ref{tc}. As can be seen, the map iterations do not agree with analytical equations exactly. This is probably due to the limited accuracy of numerical calculations, that near such critical point can play crucial role. Numerical noise can tip the system over the edge into the parallel state.\\

We have not however introduced the iterated map to investigate what we can analytically. We have performed the iterations of the map for the scale-free distribution of degrees. The distribution was generated with same algorithm of Barabasi-Albert network growth as in Monte-Carlo simulations, to allow better comparison.

Looking at Figure \ref{tode} it is evident, that the analytic results based on assumption of second order phase transition are incorrect. For small $p$, the first order phase transition critical temperature is linearly dependent on the parameter $p$, but with different factor. For higher inter-network connection number, the dependence is no longer linear.

\begin{figure}[tb]
 \vskip 0.5cm
 \centerline{\epsfig{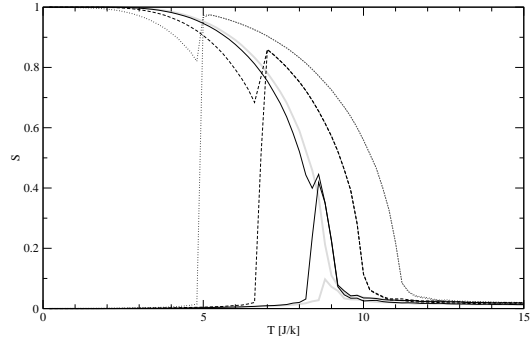}}
 \caption{\label{sdyna}Dependence of $\abs{(S_A+S_B)/2}$ (lines with near zero value below $T_c$) and $\abs{S_A}$ on the temperature for Monte-Carlo simluations. The thick grey lines are for $p=0$, solid lines are for $p=0.02$, dashed for $p=0.1$ and dotted for $p=0.2$. The simulations start from antiparallel ordering $S_A=-S_B=1$, $N_A=N_B=5000$, $\mean{k_A}=\mean{k_B}=10$ and are performed for $\tau=100$.}
\end{figure}

\section{Monte-Carlo simulations}
Our investigation would not be complete, if we didn't use numerical simulations to test our results. We have performed Monte-Carlo simulations of the Ising model on two inter-connected Barabasi-Albert networks. We have simulated networks where $N_A=N_B=5000$ and $\mean{k_A}=\mean{k_B}=10$.\\
The simulation for each temperature $T$ is independent on others and starts from an antiparallely ordered system $S_A=-S_B=1$. The dynamics of Ising model are applied for $\tau=100$ time steps and then we perform measurement of average $\abs{(S_A+S_B)/2}$ and $\abs{S_A}$ during another $\tau=100$ time steps. One time step equals $N_A+N_B$ random single node updates, what means average one update per node. We have chosen the time $\tau$ so that the network has enough time to relax to the equilibrium state, but not enough to reliably jump to the parallel state due to the temperature noise. With our chosen $\tau=100$ value, the results change little if we increase it further, thus we can be sure that the time is enough for the network to relax.\\

\begin{figure}[tb]
 \vskip 0.5cm
 \centerline{\epsfig{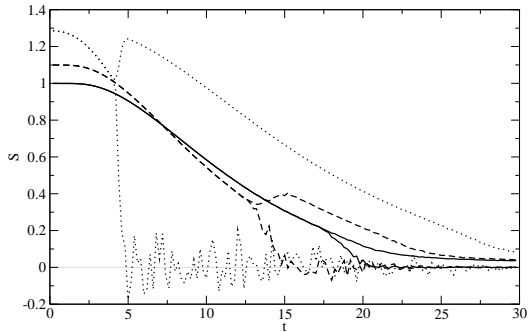}}
 \caption{\label{ba_dyna}Dependence of weighted spin absolute value $\abs{S}$ and of weighted spin of single network $S_A$. The results are for $N_A=N_B=5000$ and $\mean{k_A}=\mean{k_B}=10$. The solid lines are for $E_{AB}=0$, dashed for $E_{AB}=5000$ and dotted for $E_{AB}=15000$. The upper lines are absolute spins, while the bottom, reaching $0$ are for single network weighted spin $S_A$. The weighted spin values above $1$ result from increasing $\mean{k}$ due to interconnections.}
\end{figure}

The example of the simulation results are presented at Figure \ref{sdyna}. We measure $\abs{(S_A+S_B)/2}$ and $\abs{S_A}$ because the first order phase transition can be spotted on graphs of these values. Since we start from antiparallel ordering, $\abs{(S_A+S_B)/2}$ is close to zero below $T_c$, as both networks have same sizes and weighter spin $S$ values, only of opposite sign, so the total is close to zero. It is not exactly zero, only because of fluctuations. Since we measure the absolute value, those fluctuations do not cancel each other, but add up, resulting in non-zero total absolute value of spin.\\
When we reach critical temperature $T_c$, $\abs{(S_A+S_B)/2}$ becomes positive as the networks order parallely. The sudden change of total weighted spin means we have first order phase transition. We have to use the absolute value, since different simulations order either with positive or negative spin with equal probability, so if we didn't use average of absolute value, we would not be able to see the transition point. As the temperature grows higher, the networks order parallely with lesser and lesser value of spin and finally at temperature $T_{c+}$ they become paramagnetic. This transition is of second order. We do not investigate this transition, as it was done before \cite{isingconnect}.\\
$\abs{S_A}$ behaves similarly, except below $T_c$ it is positive. At $T_c$ there is a sudden change, since the weighted spin of networks ordered paralelly is higher at same temperature $T$ than the weighted spin of networks ordered antiparalelly.\\
We assume that the local minimum of $\abs{S_A}$ is at $T_c$, just before the networks start to order parallely. We investigate $\abs{(S_A+S_B)/2}$ only to confirm that the minimum of $\abs{S_A}$ is indeed at the temperature where the system is about to switch to parallel state and $\abs{(S_A+S_B)/2}$ becomes positive.\\
We have investigated the dependence of $T_c$ on the number of inter-network links $E_{AB} \sim p$. First, we have taken the case of $k=$ const, to test how the simulations compare to analytic results and map iterations. The results (Fig.\ref{tc}) indicate, that while critical temperatures $T_c$ are different than predicted analytically, but the error is not large. The fact, that the temperatures drop to zero at around $p=0.5$, not around $p=1$ shows that mean-field method does not describe the dynamics of the system accurately. In real systems, nodes in one network can be influenced by the second network stronger than by their own.\\
Our main results concern the case of the Barabasi-Albert networks. The weighted spin against temperature for several different interconnection densities is shown in Figure \ref{ba_dyna}.

\begin{figure}[tb]
 \vskip 0.5cm
 \centerline{\epsfig{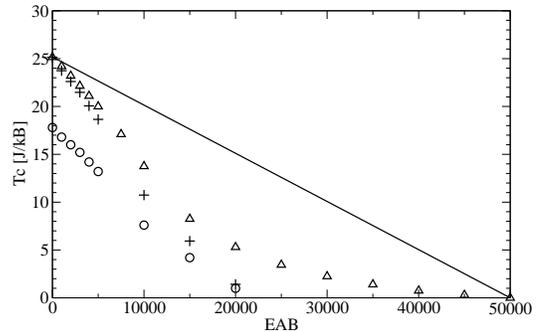}}
 \caption{\label{final}Dependence of $T_c$ on the inter-network link number $E_{AB} \sim p$. The line is analytic prediction of second order phase transition model, triangles are map iterations that display first order phase transition, while circles are data obtained from Monte-Carlo simulations. The plus symbols are same as circles, but re-scaled to have same value at $E_{AB}=0$ as map iterations.}
\end{figure}

The jumps of $\abs{S}$ at values far from zero prove that the transition is indeed of the first order as expected, since second order phase transition would have weighted spin drop to zero (or almost zero, because of fluctuations) before jumping back to positive (parallel ordering). The dependence of $T_c$ on $E_{AB}$, obtained from data that are partially shown at Figure \ref{ba_dyna} is shown at Figure \ref{final}. The critical temperature $T_c$ is much lower than predicted by either analytics or map iterations, but this is general problem with Ising model critical temperature in B-A network. If for a moment, we omit this and re-scale our results to have same value for unconnected networks, we obtain relatively good agreement with map iterations. Above $E_{AB} \approx 15000$, the results from map iterations and simulations start to differ strongly. This is possibly due to simulations having limited system size, so antiparallel ordering is quickly destroyed by fluctuations and system reverts to parallel ordering that has lower energy. However, for lower interconnection densities, the simulations agree with map iterations.

\section{Conclusions}
We have shown that in a system of two connected networks, one of two temperature driven phase transitions is of first order, unlike classical Ising phase transitions that are of second order. The dependence of the critical temperature on the interaction strength between the networks is complex. The temperatures are lower than a theory based on second order phase transition predicts. The conclusions are backed up by the numerical simulations.


\begin{acknowledgments}
K.Suchecki acknowledges the support of the EU Grant {\it Measuring and Modelling Complex Networks Across Domains} (MMCOMNET). J.Ho\l yst acknowledges the support of the EU Grant {\it Critical Events in Evolving Networks} (CREEN).
\end{acknowledgments}


\end{document}